\documentclass{article}

\usepackage{PRIMEarxiv}

\usepackage[utf8]{inputenc}
\usepackage[T1]{fontenc}
\usepackage{hyperref}
\usepackage{url}
\usepackage{booktabs}
\usepackage{amsfonts}
\usepackage{amsmath}
\usepackage{nicefrac}
\usepackage{microtype}
\usepackage{fancyhdr}
\usepackage{graphicx}
\usepackage{float}
\usepackage{multirow}
\usepackage{array}
\usepackage{natbib}
\usepackage{caption}

\graphicspath{{figures/}}

\newcolumntype{C}[1]{>{\centering\arraybackslash}p{#1}}
\newcolumntype{L}[1]{>{\raggedright\arraybackslash}p{#1}}

\newcommand{\kalshi}{\textsc{Kalshi}}

\title{Do Prediction Markets Forecast Cryptocurrency Volatility? Evidence from Kalshi Macro Contracts}

\author{
  Hardhik Mohanty\textsuperscript{*}, Bhaskar Krishnamachari \\
  Viterbi School of Engineering \\
  University of Southern California \\
  Los Angeles, CA \\
}

\begin{document}
\fancyhead[RO]{\textit{Mohanty and Krishnamachari (2026)}}
\maketitle
\renewcommand{\thefootnote}{\fnsymbol{footnote}}
\footnotetext[1]{Corresponding author. Email: \texttt{hmohanty@usc.edu}}
\renewcommand{\thefootnote}{\arabic{footnote}}
\setcounter{footnote}{0}

\begin{abstract}
Daily probability changes in \kalshi{} macro prediction markets forecast cryptocurrency realized volatility through two distinct channels. The monetary policy channel, measured by Fed rate repricing on KXFED contracts, predicts Bitcoin volatility in sample with $t = 3.63$ and $p < 0.001$ but exhibits regime dependence tied to the 2024--2025 rate-cutting cycle. The recession risk signal from KXRECSSNBER proves more stable out of sample, delivering an MSFE ratio of 0.979 with Clark--West $p = 0.020$. The inflation channel, measured by CPI repricing on KXCPI contracts, predicts altcoin volatility for Ethereum, Solana, Cardano, and Chainlink with $t$-statistics ranging from $-2.1$ to $-3.4$ and out-of-sample gains for Ethereum at MSFE $= 0.959$ with $p = 0.010$ and Solana at $p = 0.048$. Both the Bitcoin--Fed-dovish and Chainlink--CPI specifications survive Benjamini--Hochberg correction at $q = 0.05$. Orthogonalization and baseline comparisons against Fed Funds futures, Treasury yields, and the Deribit implied volatility index confirm that these signals carry information not embedded in conventional financial instruments. The sample covers ten Kalshi event series and six cryptocurrency assets over January 2023 to March 2026.
\end{abstract}

\keywords{Prediction markets \and Kalshi \and Cryptocurrency \and Realized volatility \and Macroeconomic uncertainty \and HAR model}

\section{Introduction}
\label{sec:intro}

Macroeconomic news moves asset prices, and the magnitude of that movement depends on how much information the news contains beyond what markets had already priced. For equities and bonds, a large literature connects Fed communication, CPI releases, and employment data to volatility through the lens of monetary policy surprises \citep{bernanke2005what, nakamura2018high, gurkaynak2005sensitivity}. Cryptocurrency markets present a harder case. They trade continuously across time zones, attract a heterogeneous global investor base, and lack the institutional anchors that make macro-financial linkages legible in traditional asset classes. Existing work finds that macro news affects crypto volatility but remains constrained by the fact that conventional surprise measures are available only on scheduled announcement days \citep{walther2019exogenous, macro_crypto_volatility2025}. Between announcements, there is no standard continuous measure of how macro expectations are shifting.

This paper asks whether daily probability changes in \kalshi{} prediction markets fill that gap. \kalshi{} is the first CFTC-regulated event contract exchange in the United States, offering binary options on Federal Reserve rate decisions, CPI prints, GDP growth, and NBER recession determinations. Contract prices represent risk-neutral probabilities of macro outcomes, which may diverge from physical probabilities due to risk premia, as emphasized by \citet{manski2006interpreting}, and their day-to-day changes provide a continuous measure of macro uncertainty repricing. For the present analysis, the distinction between risk-neutral and physical probabilities matters less than it does for point forecasting of macro outcomes, because the signal of interest is the magnitude of daily repricing rather than the probability level itself. Risk premia that are roughly constant across days cancel in the first difference, so the daily change largely reflects updating of beliefs rather than shifts in the risk premium. The prediction market literature has long argued that such markets aggregate dispersed information efficiently \citep{wolfers2006prediction, arrow2008promise, snowberg2013prediction}, and recent work by \citet{diercks2026kalshi} confirms that Kalshi prices contain genuine economic information for macro forecasting. We construct daily volume-weighted probability change signals for eight Kalshi series and test each against five-day ahead realized volatility for Bitcoin, Ethereum, Solana, Cardano, Avalanche, and Chainlink.

Two distinct questions motivate the analysis: First is whether Kalshi generates signals with forecasting content beyond what conventional financial instruments already provide. The orthogonalization exercises and baseline comparisons in Section~\ref{sec:robustness} address this directly by benchmarking Kalshi against Fed Funds futures, Treasury yields, and implied volatility measures. The second question is which macro channels matter for which crypto assets, and through what economic mechanism. The cross-asset results in Section~\ref{sec:results} address this.

The results are organized around these two questions. On the first, Kalshi signals are largely orthogonal to conventional macro metrics. The first-stage $R^2$ is 2.3 percent for the monetary policy signal and 7.5 percent for the CPI signal, and both retain predictive power in joint models that include Fed Funds rate changes and 10-year Treasury returns. On the second, the evidence reveals channel specificity: Bitcoin volatility responds to monetary policy repricing while altcoin volatility responds to CPI. The monetary policy channel exhibits regime dependence. The Fed-dovish signal is the strongest in-sample predictor for Bitcoin with $t = 3.63$ and $p < 0.001$, but it does not deliver out-of-sample forecast gains over the full evaluation window. The CSSED curves show that forecast gains accumulate during the 2024--2025 rate-cutting cycle and reverse once the cycle ends. The recession risk signal from KXRECSSNBER, which draws on a more slowly evolving macro state, provides the most reliable out-of-sample gains for Bitcoin with an MSFE ratio of 0.979 and Clark--West $p = 0.020$. For altcoins, the CPI channel is more stable: out-of-sample gains are confirmed for Ethereum with MSFE 0.959 and $p = 0.010$, and for Solana with MSFE 0.983 and $p = 0.048$.

To the extent that Bitcoin volatility is governed by institutional investors reacting to monetary policy expectations, and altcoin volatility by retail participants navigating inflation regime uncertainty, we expect Fed signals to load significantly on BTC and CPI signals on smaller tokens. The cross-asset results are consistent with this prediction, though the data cannot test the mechanism directly in the absence of participant-level information. The paper contributes to the crypto volatility literature \citep{katsiampa2017volatility, walther2019exogenous, liang2022which, fang2020predicting} by introducing a new class of continuous macro signal derived from regulated prediction markets, and to the prediction market literature \citep{wolfers2006prediction, manski2006interpreting, diercks2026kalshi} by providing the first evidence that event contract prices forecast crypto asset volatility at daily frequency. Section~\ref{sec:data} describes the data and signal construction. Section~\ref{sec:method} presents the empirical framework. Section~\ref{sec:results} reports in-sample results, while Section~\ref{sec:oos} examines out-of-sample forecasting performance. Section~\ref{sec:robustness} presents robustness checks, Section~\ref{sec:implications} discusses practical implications, and Section~\ref{sec:conclusion} concludes.

\section{Data}
\label{sec:data}

\subsection{Kalshi Prediction Markets}

\kalshi{} launched in 2021 as the first CFTC-regulated event contract exchange in the United States. Each contract is a binary option on a macro outcome with a payoff of \$1 if the outcome occurs and \$0 otherwise. The closing price on any day represents the market's risk-neutral probability of the outcome.
We collect daily closing prices, trading volumes, and open interest for ten contract series via the public Kalshi API, covering January 2023 through March 2026. Table~\ref{tab:kalshi_coverage} in the appendix lists each series, its macro domain, and the number of usable observations. Coverage varies: KXRECSSNBER (NBER recession risk) has 569 usable days while KXPCECORE (core PCE) has 294, and KXUSNFP (non-farm payrolls) has zero, reflecting the irregular listing and expiration of event-specific contracts.

Within each series $s$ on day $t$, multiple contracts with different outcome thresholds trade simultaneously. We aggregate them using volume-weighted averaging:
\begin{equation}
  \Delta^{vw}_{s,t} = \frac{\sum_{j \in \mathcal{J}_s} V_{j,t} \cdot \Delta p_{j,t}}{\sum_{j \in \mathcal{J}_s} V_{j,t}},
  \label{eq:vwdelta}
\end{equation}
where $\mathcal{J}_s$ is the set of active contracts in series $s$, $V_{j,t}$ is dollar volume, and $\Delta p_{j,t}$ is the closing-to-closing probability change. The absolute value $|\Delta^{vw}_{s,t}|$ measures the magnitude of macro repricing regardless of direction and serves as the primary signal. KXRATECUT (28 observations) and KXUSNFP are excluded due to insufficient coverage.

Contracts trade on a central limit order book with a minimum tick of \$0.01. Near the midpoint of the price distribution ($\sim$\$0.50), the implied round-trip spread is roughly two cents, widening to five cents at the tails. Dollar trading volume on KXFED has a sample median of \$9,326, with single-day peaks exceeding \$651,000. Open interest is larger and more persistent: \$278,805 at the median for KXFED, \$147,879 for KXCPI, and \$456,648 for KXRECSSNBER, the highest among all series. Volume concentrates around scheduled announcement dates (FOMC meetings for KXFED, BLS releases for KXCPI). The KXRECSSNBER pattern differs: low daily turnover paired with high open interest suggests a continuous-tenor contract held for hedging rather than event-driven speculation. The public API does not provide individual contract bid-ask quotes, so execution costs cannot be directly measured from the data.

The primary signal is the absolute volume-weighted change. We also evaluate alternatives: a five-day exponential moving average of the lagged signal, and a cross-sectional composite averaging absolute signals across all active series on a given day. The EMA is substantially attenuated relative to the point-in-time signal ($t = 1.24$ for BTC versus 3.63), and the composite yields near-zero $t$-statistics ($-0.79$ for BTC, $-0.59$ for LINK). Predictive information resides in discrete repricing events within specific macro series, not in aggregate trend.
For monetary policy, we construct a directional Fed-dovish signal: $-\Delta^{vw}_{\text{KXFED},t}$, positive when rate expectations shift downward. Analogous directional signals are constructed for CPI and NFP.

\subsection{Cryptocurrency Returns and Realized Volatility}

Daily closing prices for Bitcoin, Ethereum, Solana, Cardano, Avalanche, and Chainlink are obtained from CoinGecko. Log returns are $r_{a,t} = \ln(P_{a,t}/P_{a,t-1})$.
The dependent variable is five-day forward realized volatility:
\begin{equation}
  \text{RVol}_{a,t}^{h=5} = \sqrt{252} \cdot \hat{\sigma}\!\left(r_{a,t+1}, \ldots, r_{a,t+5}\right),
  \label{eq:rvol}
\end{equation}
where $\hat\sigma$ denotes sample standard deviation.\footnote{We use $\sqrt{252}$ rather than $\sqrt{365}$ for comparability with the equity volatility literature. Rescaling to calendar days would multiply all reported values by $\sqrt{365/252} \approx 1.20$ but leave the qualitative results unchanged.} We also use one-day absolute return $|r_{a,t+1}|$ for the non-overlapping robustness check. Table~\ref{tab:descriptive} in the appendix reports descriptive statistics. Bitcoin has the lowest mean annualized five-day realized volatility (0.634), consistent with deeper liquidity and higher institutional participation. Solana and Avalanche exceed 0.90.

\subsection{Market Controls}

All regressions include the CBOE VIX closing level, the DXY index daily return, and the S\&P 500 daily return as controls. These absorb common risk-off dynamics that could cause Kalshi signals to serve as a metric for broad market stress. The daily change in implied Fed Funds futures rate enters as a conventional policy benchmark in orthogonalization and baseline comparison exercises.

Kalshi closing prices are recorded at 4:00 PM ET; cryptocurrency prices use CoinGecko's midnight-UTC close (7:00 PM ET) for the same calendar date. Because the dependent variable begins at $t+1$, a minimum gap of roughly 21 hours separates the signal from the forecast window. VIX, S\&P 500, and DXY controls are aligned to the 4:00 PM ET close. Weekends and U.S. holidays are excluded, dropping approximately 30 percent of calendar days but ensuring all regressors are observed before the forecast window opens. The merged dataset spans January 2023 to March 2026 ($N = 1{,}183$ calendar days). Effective sample sizes after merging with each Kalshi series range from 193 to 569 observations.

\section{Empirical Strategy}
\label{sec:method}

\subsection{HAR Benchmark}

The Heterogeneous Autoregressive (HAR) model of \citet{corsi2009simple} is the standard benchmark for realized volatility forecasting, decomposing persistence across daily, weekly, and monthly horizons:
\begin{equation}
  \text{RVol}_{a,t}^{h=5} = \alpha + \beta_1 |r_{a,t-1}| + \beta_2 \bar{|r|}_{a,t-1}^{(5)} + \beta_3 \bar{|r|}_{a,t-1}^{(20)} + \varepsilon_{a,t},
  \tag{M1}
  \label{eq:m1}
\end{equation}
where $\bar{|r|}^{(k)}_{a,t-1}$ is the lagged $k$-day rolling mean of absolute daily log returns.

\subsection{Augmented Models}

Adding market controls and the Kalshi signal sequentially gives:
\begin{align}
  \text{RVol}_{a,t}^{h=5} &= \alpha + \boldsymbol{\beta}' \mathbf{HAR}_{a,t} + \boldsymbol{\gamma}' \mathbf{Ctrl}_t + \varepsilon_{a,t}, \tag{M2} \label{eq:m2} \\
  \text{RVol}_{a,t}^{h=5} &= \alpha + \boldsymbol{\beta}' \mathbf{HAR}_{a,t} + \boldsymbol{\gamma}' \mathbf{Ctrl}_t + \delta \cdot \text{Kalshi}_{s,t-1} + \varepsilon_{a,t}, \tag{M3} \label{eq:m3}
\end{align}
where $\mathbf{Ctrl}_t = (\text{VIX}_t, r_{\text{DXY},t}, r_{\text{SP500},t})'$ and $\text{Kalshi}_{s,t-1}$ is the one-day lagged signal. The lag ensures the signal is publicly observable before the forecast window opens.

\subsection{Inference and Multiple Testing}

Five-day realized volatility creates overlapping observations with MA($h-1$) serial correlation. We use Newey--West HAC standard errors with 5 lags throughout \citep{newey1987simple}. For non-overlapping specifications, HC3 standard errors apply.
With 60 signal-by-asset combinations in the cross-sectional analysis, multiple testing is a genuine concern \citep{harvey2016and, white2000reality}. We apply the Benjamini--Hochberg FDR procedure at $q = 0.05$ \citep{benjamini1995controlling}.

\subsection{Out-of-Sample Evaluation}

We use an expanding-window scheme with an initial training window of 120 days. Forecast accuracy is assessed using the \citet{clark2007approximately} statistic for nested model comparison:
\begin{equation}
  \hat{f}_t = e_{b,t}^2 - \left[ e_{a,t}^2 - (\hat{y}_{b,t} - \hat{y}_{a,t})^2 \right],
  \label{eq:cw}
\end{equation}
where $e_{b,t}$ and $e_{a,t}$ are baseline (M2) and augmented (M3) forecast errors. The CW $t$-statistic is tested one-sided. We also report the MSFE ratio and OOS $R^2$ relative to the expanding historical mean.

\section{In-Sample Results}
\label{sec:results}

\subsection{Cross-Asset Heterogeneity}
\begin{figure}[!h]
  \centering
  \includegraphics[width=\textwidth]{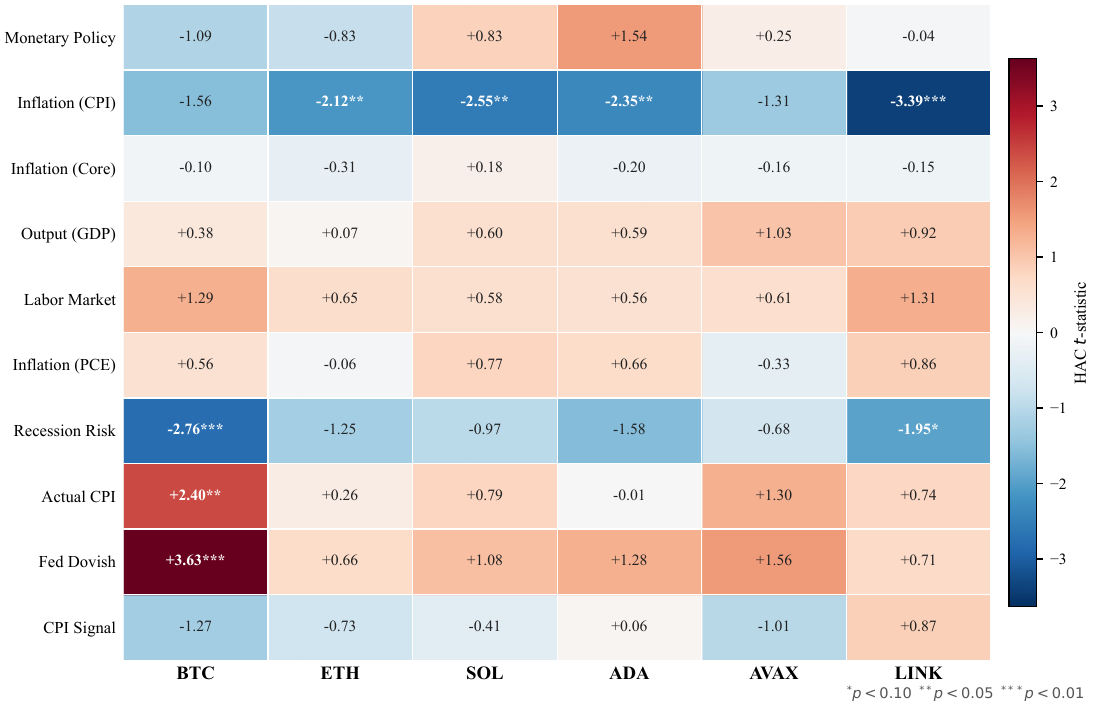}
  \caption{HAC $t$-statistics from model M3 for each signal-by-asset pair. Dependent variable: 5-day ahead annualized realized volatility. Newey--West standard errors with 5 lags. $^* p < 0.10$, $^{**} p < 0.05$, $^{***} p < 0.01$. Dashes indicate the series was inactive for that estimation window.}
  \label{fig:heatmap}
\end{figure}

Figure~\ref{fig:heatmap} presents the central empirical pattern: HAC $t$-statistics from model M3 across all 10 Kalshi signals and 6 cryptocurrency assets.
The dominant pattern is channel specificity. The Fed-dovish signal predicts Bitcoin volatility ($t = 3.63$, $p < 0.001$) but shows no significant relationship with any altcoin. The CPI signal (KXCPI) is negative and significant for ETH, SOL, ADA, and LINK ($t = -2.1$ to $-3.4$) but not for BTC or AVAX. Recession risk (KXRECSSNBER) is significant for BTC ($t = -2.76$) but not for other assets. The remaining series show mostly null results, consistent with either insufficient liquidity or redundancy.

This pattern is inconsistent with a simple risk-off story, in which all crypto assets would load similarly on any macro uncertainty signal. Instead, different macro transmission channels appear to operate for different assets, a finding consistent with heterogeneous investor compositions across the cryptocurrency market.

\subsection{Bitcoin and Monetary Policy}

\begin{table}[!ht]
  \centering
  \caption{Nested model comparison for Bitcoin five-day realized volatility. HAC standard errors (Newey--West, 5 lags) in parentheses. $N = 318$.}
  \label{tab:insample_btc}
  \small
  \begin{tabular}{lC{2.2cm}C{2.2cm}C{2.2cm}}
    \toprule
                      & M1            & M2            & M3            \\
    \midrule
    HAR (1-day lag)   & 0.581         & 0.474         & 0.395         \\
                      & (0.640)       & (0.589)       & (0.560)       \\[3pt]
    HAR (5-day lag)   & 6.047$^{*}$   & 6.075$^{*}$   & 6.388$^{*}$   \\
                      & (3.179)       & (3.415)       & (3.348)       \\[3pt]
    HAR (20-day lag)  & 0.351         & $-$1.086      & $-$1.323      \\
                      & (3.285)       & (3.351)       & (3.318)       \\[3pt]
    VIX               &               & 0.005         & 0.005         \\
                      &               & (0.004)       & (0.004)       \\[3pt]
    DXY return        &               & 4.885$^{*}$   & 4.646$^{*}$   \\
                      &               & (2.896)       & (2.823)       \\[3pt]
    S\&P 500 return   &               & $-$2.859$^{**}$ & $-$2.770$^{**}$ \\
                      &               & (1.234)       & (1.217)       \\[3pt]
    Fed-dovish signal &               &               & 0.639$^{***}$ \\
                      &               &               & (0.176)       \\[3pt]
    \midrule
    Adj.\ $R^2$       & 0.093         & 0.141         & 0.155         \\
    $N$               & 318           & 318           & 318           \\
    \bottomrule
  \end{tabular}
  \par\smallskip
  \footnotesize $^* p < 0.10$, $^{**} p < 0.05$, $^{***} p < 0.01$. Constant included but not reported.
\end{table}

Table~\ref{tab:insample_btc} reports nested model comparisons for Bitcoin. The HAR benchmark achieves an adjusted $R^2$ of 9.3 percent (M1). Adding market controls raises the adjusted $R^2$ to 14.1 percent (M2). The Fed-dovish signal further increases adjusted $R^2$ to 15.5 percent, with $\hat\delta = 0.639$ and $t = 3.63$.
The interquartile range of the Fed-dovish signal is approximately 0.023, so a shift from the 25th to the 75th percentile raises predicted five-day realized volatility by $0.639 \times 0.023 = 0.015$ annualized units, roughly 2.4 percent of the full-sample mean Bitcoin realized volatility of 0.634.

The signal is defined as $-\Delta^{vw}_{\text{KXFED},t-1}$, positive when rate expectations shift downward. The finding that dovish surprises predict higher Bitcoin volatility over the following week admits several interpretations: Dovish shifts may signal macro weakness, as unexpected Fed softening typically accompanies deteriorating economic data, generating uncertainty that elevates volatility across risk-sensitive assets. Alternatively, Bitcoin pricing is known to respond to dollar liquidity conditions \citep{demir2018does, bouri2017hedge}, and shifts in rate expectations may trigger position rebalancing that propagates through the following week. A third possibility is that Kalshi and crypto markets share overlapping participant pools whose information sets are correlated, producing statistical association without a direct transmission channel. Our lead-lag test (Section~\ref{sec:robustness}) rules out the simplest form of reverse causality but cannot discriminate among these three forward-looking mechanisms. All three interpretations are plausible and not mutually exclusive.

\subsection{Altcoins and Inflation}

\begin{table}[H]
  \centering
  \caption{CPI prediction market signal (KXCPI, absolute probability change, lagged one day) in model M3. Dependent variable: five-day ahead realized volatility. HAC standard errors (Newey--West, 5 lags). $N = 264$.}
  \label{tab:insample_cpi}
  \small
  \begin{tabular}{lC{2.2cm}C{2.0cm}C{1.8cm}C{1.8cm}}
    \toprule
    Asset     & Coef.              & HAC $t$ & $p$-value & Adj. $R^2$ \\
    \midrule
    Bitcoin   & $-$0.454           & $-$1.56 & 0.119 & 0.160 \\
    Ethereum  & $-$1.209$^{**}$    & $-$2.12 & 0.034 & 0.108 \\
    Solana    & $-$0.850$^{**}$    & $-$2.55 & 0.011 & 0.083 \\
    Cardano   & $-$0.982$^{**}$    & $-$2.35 & 0.019 & 0.082 \\
    Avalanche & $-$0.834           & $-$1.31 & 0.191 & 0.017 \\
    Chainlink & $-$1.262$^{***}$   & $-$3.39 & 0.001 & 0.110 \\
    \bottomrule
  \end{tabular}
  \par\smallskip
  \footnotesize $^* p < 0.10$, $^{**} p < 0.05$, $^{***} p < 0.01$. All models include HAR components and market controls.
\end{table}

Table~\ref{tab:insample_cpi} reports CPI signal results across all six assets. The negative coefficient indicates that larger absolute CPI probability shifts predict lower next-week realized volatility. The natural interpretation is uncertainty resolution: large Kalshi CPI moves concentrate on and around scheduled BLS release dates, and the resolution of the inflation data event reduces volatility that had been building in anticipation. This channel operates more strongly for altcoins than for Bitcoin, consistent with the view that altcoin investor bases are more exposed to inflation regime uncertainty than to monetary policy path uncertainty \citep{corbet2018exploring, liu2022risks}.

Two rival accounts of the negative coefficient must be distinguished. The first is genuine uncertainty resolution: Kalshi CPI moves capture the updating of inflation expectations, and the resolution of that uncertainty mechanically reduces the component of altcoin volatility attributable to CPI ambiguity. The second is a calendar effect: CPI release days are followed by structurally lower altcoin volatility for unrelated reasons, and Kalshi CPI moves are large on those days by construction. To discriminate, we augment model M3 with an indicator for the three-day window around each BLS CPI release date. When this release-window dummy is included, the KXCPI coefficient retains its sign and significance for all three altcoins: Chainlink ($t = -3.21$, $p = 0.001$), Solana ($t = -2.63$, $p = 0.009$), and Ethereum ($t = -2.07$, $p = 0.039$). The release-window dummy itself is insignificant for Ethereum and Solana and only modestly significant for Chainlink ($t = -2.16$). As a further check, we restrict the sample to non-release days only, excluding the three-day window around each BLS date. Solana ($t = -2.80$) and Chainlink ($t = -3.07$) remain strongly significant, while Ethereum is marginal ($t = -1.81$, $p = 0.071$). The CPI signal operates predominantly through uncertainty resolution rather than a mechanical calendar effect, though the partial attenuation for Ethereum on non-release days suggests some contribution from the release calendar.

Avalanche is the exception among altcoins: the CPI coefficient is negative ($-0.834$) but insignificant ($t = -1.31$). The M3 adjusted $R^2$ for AVAX is 0.017, far below the 0.082 to 0.110 range for the significant altcoins, indicating that the HAR-plus-controls framework explains very little AVAX variance. AVAX had the highest idiosyncratic volatility in the sample alongside distinct ecosystem dynamics (institutional subnet activity, concentrated DeFi exposure), and the macro signal is likely overwhelmed by asset-specific noise.

\subsection{Horizon Profile}

\begin{figure}[H]
  \centering
  \includegraphics[width=\textwidth]{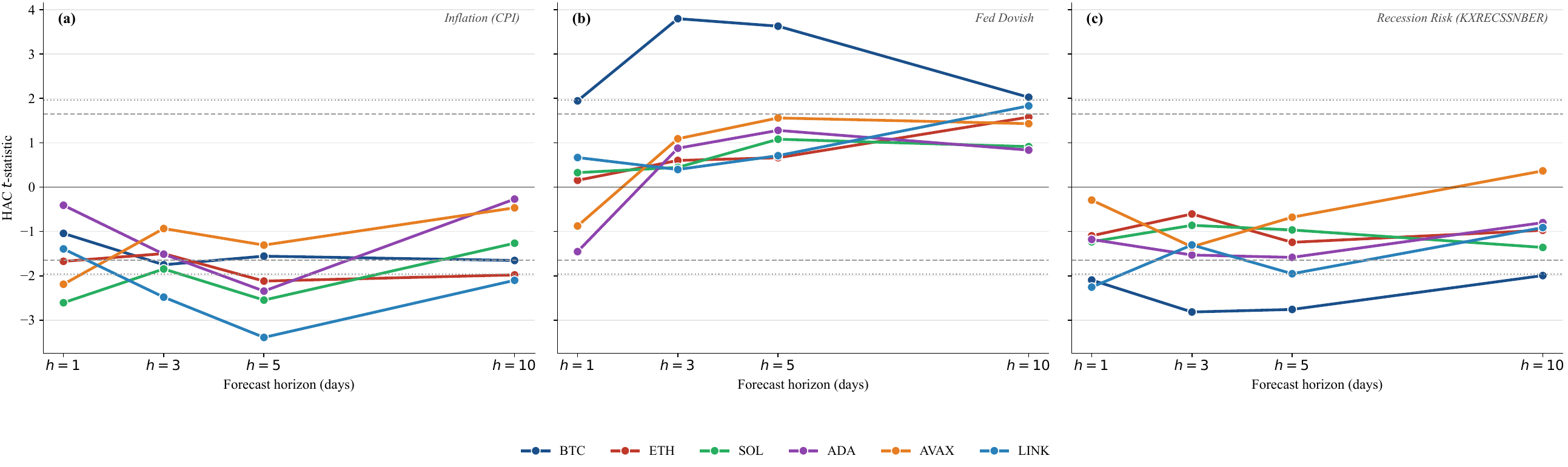}
  \caption{HAC $t$-statistics for model M3 across forecast horizons $h = 1, 3, 5, 10$ days. Dashed lines at $|t| = 1.645$ (10\%) and $|t| = 1.960$ (5\%). HAC lags $= \min(h, 5)$.}
  \label{fig:horizon}
\end{figure}

Figure~\ref{fig:horizon} plots HAC $t$-statistics across four forecast horizons ($h = 1, 3, 5, 10$ days). The Fed-dovish signal for Bitcoin peaks at $h = 3$ to $h = 5$ and fades at both shorter and longer horizons. This profile is consistent with a multi-day diffusion of monetary policy uncertainty into realized volatility rather than an instantaneous response. The CPI signal for LINK and SOL concentrates at $h = 5$, while ETH shows more persistent effects through $h = 10$.

If prediction market repricing were merely correlated with contemporaneous volatility, the effect would weaken at longer horizons. The five-day peak is consistent with a genuine predictive relationship rather than contemporaneous correlation.

\subsection{Best Signal by Asset}

\begin{figure}[H]
  \centering
  \includegraphics[width=0.85\textwidth]{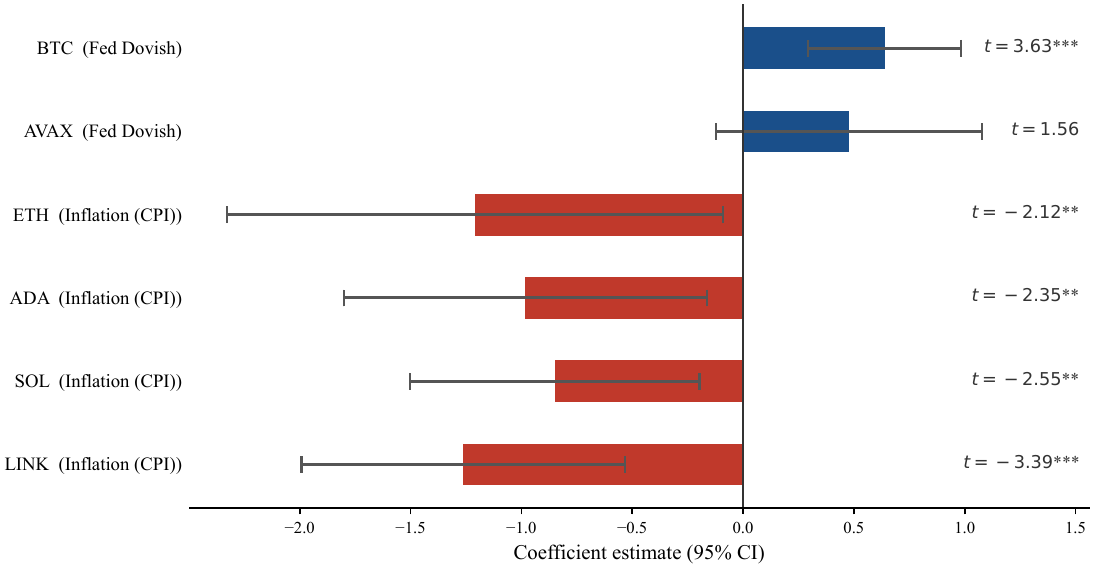}
  \caption{Coefficient estimates with 95\% confidence intervals for the best-performing Kalshi signal per cryptocurrency (five-day realized volatility, model M3). HAC standard errors (Newey--West, 5 lags).}
  \label{fig:forest}
\end{figure}

Figure~\ref{fig:forest} summarizes the strongest Kalshi signal for each cryptocurrency at the five-day horizon. Because signal selection is conducted within the full sample, these estimates should be interpreted as characterizing the dominant macro channel for each asset rather than as unconditional forecasting rules. The Benjamini--Hochberg correction in Section~\ref{sec:robustness}, applied across all 60 tested specifications, provides the appropriate inferential benchmark. Bitcoin and Avalanche respond to the monetary policy channel (positive coefficients), while Ethereum, Solana, Cardano, and Chainlink respond to CPI repricing (negative coefficients). Chainlink shows the largest absolute $t$-statistic among the altcoins.

\section{Out-of-Sample Evaluation}
\label{sec:oos}

Table~\ref{tab:oos} summarizes expanding-window out-of-sample results. An MSFE ratio below 1.0 indicates the augmented model outperforms the baseline. The Clark--West $p$-value tests the one-sided null that the Kalshi signal adds nothing.

\begin{table}[H]
  \centering
  \caption{Out-of-sample forecast evaluation (expanding window, initial period = 120 obs.). MSFE ratio: M3 to M2 mean squared forecast error. Best signal per asset in bold.}
  \label{tab:oos}
  \small
  \begin{tabular}{llC{1.6cm}C{1.6cm}C{1.6cm}C{1.6cm}}
    \toprule
    Asset & Signal & $N_{\text{oos}}$ & OOS $R^2$ & MSFE ratio & CW $p$ \\
    \midrule
    \multirow{3}{*}{Bitcoin}
      & Fed Dovish               & 198 & 0.118 & 1.009 & 0.446 \\
      & \textbf{Recession Risk}  & \textbf{269} & \textbf{0.126} & \textbf{0.979} & \textbf{0.020}$^{**}$ \\
      & Inflation (CPI)          & 144 & 0.095 & 0.995 & 0.231 \\[4pt]
    \multirow{3}{*}{Ethereum}
      & Fed Dovish               & 198 & $-$0.042 & 1.004 & 0.893 \\
      & \textbf{Inflation (CPI)} & \textbf{144} & \textbf{0.068} & \textbf{0.959} & \textbf{0.010}$^{**}$ \\
      & Recession Risk           & 269 & 0.051 & 1.000 & 0.421 \\[4pt]
    \multirow{2}{*}{Solana}
      & \textbf{Inflation (CPI)} & \textbf{144} & \textbf{0.048} & \textbf{0.983} & \textbf{0.048}$^{**}$ \\
      & Monetary Policy          & 198 & 0.012 & 0.999 & 0.325 \\[4pt]
    Cardano   & \textbf{Monetary Policy} & \textbf{198} & \textbf{0.011} & \textbf{0.992} & \textbf{0.041}$^{**}$ \\[4pt]
    Avalanche & Inflation (CPI)          & 144 & $-$0.045 & 0.996 & 0.171 \\[4pt]
    Chainlink & Inflation (CPI)          & 144 & 0.060 & 0.992 & 0.121 \\
    \bottomrule
  \end{tabular}
  \par\smallskip
  \footnotesize $^* p < 0.10$, $^{**} p < 0.05$, $^{***} p < 0.01$.
\end{table}

Four asset-signal pairs achieve MSFE below 1.0 with significant Clark--West statistics: BTC with Recession Risk, ETH and SOL with CPI, and ADA with Monetary Policy. These gains, while modest in absolute terms (2 to 4 percent MSFE improvement), are notable because OOS improvements in volatile asset classes like crypto are rare, and the CW critical values are conservative.

\subsection{Regime Dependence of the Monetary Policy Channel}

The most important feature of Table~\ref{tab:oos} is the divergence between the Fed-dovish signal's in-sample strength ($t = 3.63$) and its out-of-sample failure (MSFE $= 1.009$). This divergence is not a puzzle but a finding about regime dependence. The Fed-dovish signal data begins in September 2024, coinciding with the start of the Fed's rate-cutting cycle, and extends through mid-2025 when the cycle concludes. The signal captures the repricing of the monetary policy path during a period when rate expectations were actively shifting, but this same concentration in a single regime limits its out-of-sample stability.

\begin{figure}[H]
  \centering
  \includegraphics[width=\textwidth]{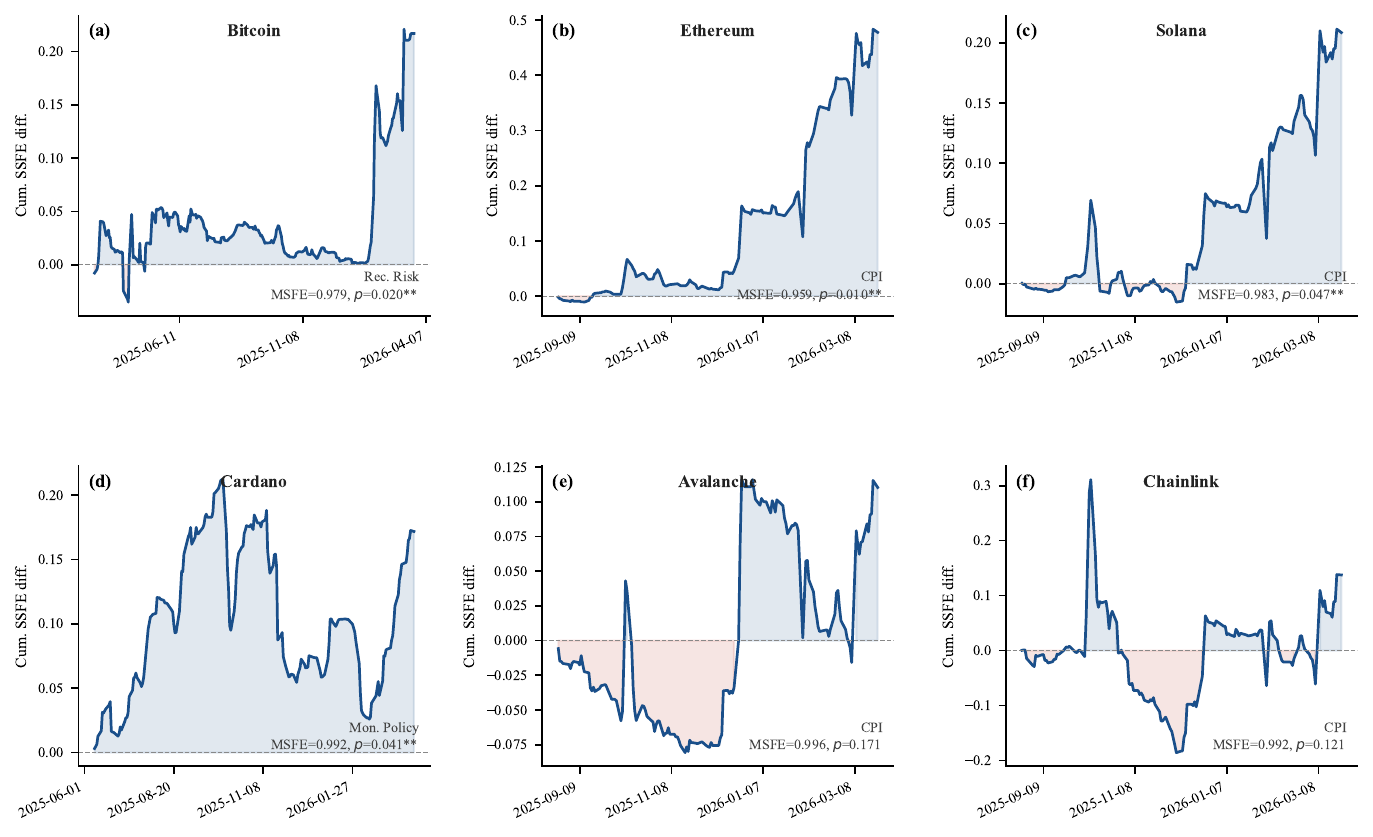}
  \caption{Cumulative sum of squared forecast error differences ($\sum_t (e_{b,t}^2 - e_{a,t}^2)$) for the best Kalshi signal per cryptocurrency. Positive values indicate the augmented model outperforms the baseline.}
  \label{fig:cssed}
\end{figure}

The CSSED curves in Figure~\ref{fig:cssed} show that forecast gains for the best-performing specifications accumulate gradually rather than in a single episode. For Bitcoin, the recession-risk signal trends upward throughout the evaluation window, consistent with a slowly evolving macro state whose information content is not tied to a single regime. The CPI signal for Ethereum and Solana similarly exhibits persistent upward drift. The Fed-dovish signal, by contrast, generates early forecast gains that subsequently reverse as the rate-cutting cycle concludes and Kalshi repricing of the policy path becomes less informative.

The practical implication is that the monetary policy channel is useful specifically during episodes of active rate repricing, not as a permanent forecasting input. Longer samples spanning multiple tightening and easing cycles will be needed to determine whether the pattern replicates. Recession risk and CPI signals, which load on more persistent macro states, offer more reliable out-of-sample performance across the evaluation window.

\section{Robustness}
\label{sec:robustness}

\subsection{Multiple Testing Correction}

Applying Benjamini--Hochberg at $q = 0.05$ across the 60 testable specifications, two survive correction: Bitcoin--Fed-dovish (adjusted $p = 0.020$) and Chainlink--CPI (adjusted $p = 0.042$). Bitcoin--Recession Risk (raw $p = 0.006$) and Solana--CPI (raw $p = 0.011$) are significant at conventional levels but fall just outside the BH threshold at $q = 0.05$ \citep{harvey2016and, white2000reality}.

\subsection{Non-Overlapping Windows}

Repeating the primary specifications using non-overlapping five-day windows eliminates overlap-induced autocorrelation and allows HC3 standard errors. The Fed-dovish coefficient for Bitcoin retains its positive sign ($t = 1.17$, $p = 0.24$) but is no longer significant, as expected given the severe reduction in sample size from 318 to 64 observations. The CPI coefficient for Chainlink yields $t = -0.71$ ($N = 54$). The loss of power with non-overlapping windows is a mechanical consequence of reducing the effective sample by a factor of five, rather than evidence against the underlying relationship.

\subsection{Orthogonalization}

We project the Fed-dovish signal onto the contemporaneous Fed Funds implied rate change, its absolute value, VIX, DXY return, and S\&P 500 return. The first-stage $R^2$ is 2.3 percent, so 97.7 percent of daily Kalshi variation is orthogonal to these observables. The residual predicts Bitcoin five-day realized volatility with $t = 3.62$ ($p < 0.001$), nearly identical to the raw signal.

For the CPI channel, we project the KXCPI signal onto the 10-year Treasury daily return (the natural bond-market metric for inflation expectations), VIX, DXY return, and S\&P 500 return. The first-stage $R^2$ is 7.5 percent. Residuals retain significant predictive power for altcoin volatility: $t = -2.29$ for Ethereum ($p = 0.022$), $t = -2.79$ for Solana ($p = 0.005$), $t = -2.60$ for Cardano ($p = 0.009$), and $t = -3.73$ for Chainlink ($p < 0.001$). The negligible attenuation confirms that Kalshi CPI markets capture information not already reflected in Treasury yields or equity implied volatility.

\subsection{Block Bootstrap}

A moving-block bootstrap with block length 5 and 2,000 resamples yields a bootstrap $p$-value of 0.035 for the Bitcoin specification. The bootstrap $p$-value is larger than the analytic HAC $p$-value of 0.0003, which is expected: block bootstrap inference is conservative relative to HAC in finite samples with strong serial dependence. The result confirms the Fed-dovish signal is significant under both inference approaches.

\subsection{Lead--Lag Test}

We compare a model using the lagged Fed-dovish signal ($t-1$) against a placebo model using the lead ($t+1$). The lagged coefficient has $t = 3.71$ ($p < 0.001$) while the lead yields $t = -0.17$ ($p = 0.87$). The absence of forward-looking predictive power is inconsistent with reverse causality.

For the CPI channel, the lead-lag test is less decisive. The lagged KXCPI signal is significant for all four altcoins as reported in Table~\ref{tab:insample_cpi}. The lead signal, however, is also significant for Ethereum ($t = -2.53$, $p = 0.012$) and Solana ($t = -3.76$, $p < 0.001$), though it attenuates for Chainlink ($t = -1.95$, $p = 0.051$). The most likely explanation is that CPI probability repricing is persistent across multiple days around BLS release dates, so both the lag and the lead capture parts of the same multi-day repricing event. This does not imply reverse causality from crypto volatility to Kalshi CPI markets, but it does mean that the timing of information flow is less sharply identified for the inflation channel than for the monetary policy channel.

\subsection{Baseline Comparison}

A key question is whether Kalshi signals merely replicate information already available from standard financial instruments. We estimate three alternative specifications for each primary channel, substituting the lagged daily Fed Funds implied rate change (M\textsubscript{FF}), the lagged 10-year Treasury return (M\textsubscript{UST}), or both jointly alongside the Kalshi signal.
For the BTC channel, M\textsubscript{FF} yields $t = -0.83$ and M\textsubscript{UST} yields $t = +0.01$, both insignificant. In the joint model, the Kalshi Fed-dovish signal retains $t = +3.45$ ($p < 0.001$) while the conventional instruments remain null. For the ETH/CPI channel, the Treasury return alone gives $t = -0.25$. In the joint model, the Kalshi CPI coefficient is $t = -2.10$ ($p = 0.035$), essentially unchanged by the inclusion of the bond return. This suggests that prediction market signals are not redundant with the information embedded in Treasury markets.

We also test against the Deribit Bitcoin Volatility Index (DVOL), a crypto-specific implied volatility measure derived from BTC and ETH options. In a joint model including both the lagged DVOL level and the Kalshi signal, the Fed-dovish coefficient for Bitcoin retains $t = +3.46$ ($p = 0.001$) while DVOL is insignificant ($t = +1.58$, $p = 0.12$). For the ETH/CPI channel, the Kalshi CPI coefficient yields $t = -2.08$ ($p = 0.037$) while ETH DVOL is again insignificant ($t = -1.52$, $p = 0.13$). The prediction market signal dominates the crypto options market's own implied volatility measure in both channels, indicating that Kalshi captures macro-driven information not reflected in the options-implied volatility surface.

\subsection{Alternative Volatility Measures}

Three alternative dependent variables test the robustness of the primary findings. GARCH(1,1) conditional variance: the Fed-dovish coefficient yields $t = +2.91$ ($p = 0.004$) for BTC, confirming the result under a parametric volatility specification. The CPI signal for ETH is insignificant under GARCH ($t = -0.69$), likely because GARCH conditional variance is a backward-looking measure that responds slowly to forward-looking signals. Log-realized volatility, $\log(\text{RVol}_{5d} + 0.001)$: $t = +3.53$ for BTC, $t = -2.41$ for ETH under the CPI signal, both consistent with the baseline results. Twenty-one-day realized volatility: the BTC effect fades ($t = +0.93$, $p = 0.35$), consistent with the weekly horizon profile documented in Section~\ref{sec:results}. The ETH 21-day result is marginal ($t = -1.78$, $p = 0.075$). The primary findings are robust to the choice of realized volatility measure at the weekly horizon but do not extend to GARCH conditional variance for the CPI channel.

\subsection{Limitations}

A few limitations warrant explicit discussion. The sample covers January 2023 to March 2026, a period that includes an inflation episode, a rate-cutting cycle, and elevated macro uncertainty. The regime dependence documented in Section~\ref{sec:oos} is itself evidence that the channels identified here may not generalize to calmer environments.
Kalshi trading volume grew substantially over the sample. While the expanding-window OOS tests show consistent forecast gains through the end of the sample, the signal's predictive content during the early, thinner-liquidity period may not reflect a permanent structural feature of the market.
The OOS evaluation uses an expanding window with a fixed initial period. Rolling-window alternatives could reveal additional regime dependence that the expanding window smooths over.

\section{Practical Implications}
\label{sec:implications}

\subsection{Volatility-Managed Portfolios}

Forecasts of realized volatility have direct applications in position sizing. In the framework of \citet{moreira2017volatility}, the portfolio weight is set inversely proportional to predicted volatility: $w_t = \bar{\sigma} / \hat{\sigma}_{t+1}$. When the Fed-dovish signal is at its 90th percentile, the augmented model predicts Bitcoin five-day realized volatility of approximately 0.348, compared to 0.324 under the HAR benchmark at mean signal values. This difference implies a roughly 7 percent reduction in the Bitcoin position weight at high-signal episodes. The opposite applies when the signal is low. While the position-level adjustment is modest, it compounds across multiple rebalancing periods, and the gains are concentrated in precisely the high-uncertainty episodes when accurate volatility forecasting matters most for risk management.

\subsection{Options Pricing}

The CPI channel has implications for altcoin options pricing. Large KXCPI moves are associated with lower next-week ETH realized volatility. If options markets set implied volatility from recent realized volatility without conditioning on prediction market repricing, they will systematically overprice straddles in the week following large CPI events. The ETH coefficient of $-1.209$ applied to a 90th-percentile signal (approximately 0.070) implies a reduction in predicted realized volatility of roughly 0.084 annualized units, or about 16 percent below the HAR benchmark prediction of 0.52. This gap is large enough to motivate short-straddle positions around CPI releases, though transaction costs in altcoin options markets would absorb a meaningful share of the expected gain.



\section{Conclusion}
\label{sec:conclusion}

Daily probability repricing in Kalshi macro prediction markets contains information about future cryptocurrency realized volatility that is not captured by the HAR benchmark, standard market controls, or conventional financial instruments including Fed Funds futures and Treasury yields. Two channels operate. The monetary policy channel, in which Fed rate repricing on Kalshi predicts Bitcoin volatility, is the strongest in-sample finding but exhibits regime dependence tied to the 2024--2025 rate-cutting cycle. The recession risk signal, which reflects a more slowly evolving macro state, provides the most reliable out-of-sample gains for Bitcoin. The inflation channel, in which CPI probability repricing predicts lower next-week altcoin volatility, is both statistically robust and more stable out of sample.

The channel specificity is consistent with Bitcoin's greater sensitivity to dollar liquidity and rate expectations relative to smaller tokens whose volatility is tied more closely to inflation regime uncertainty. A single macro uncertainty metric cannot adequately capture the heterogeneous macro sensitivities across cryptocurrency markets. Investors and risk managers seeking to forecast crypto volatility can improve on standard models by monitoring prediction market signals matched to each asset's dominant macro channel, with the caveat that the monetary policy channel's value is concentrated in episodes of active rate repricing.

Several directions for future research follow directly: Intraday data would allow tighter identification of the timing of information transmission from prediction markets to crypto volatility. Participant-level data from Kalshi, if it became available, would enable direct tests of whether the same traders operate in both markets. Cross-platform validation against other event markets (Polymarket, PredictIt) would test generalizability beyond Kalshi. Longer samples spanning multiple monetary policy regimes will clarify whether the regime dependence documented here is a feature of our specific sample or a permanent characteristic of the monetary policy channel.

\bibliographystyle{plainnat}
\bibliography{references}

@techreport{wolfers2006prediction,
  author      = {Wolfers, Justin and Zitzewitz, Eric},
  title       = {Prediction markets in theory and practice},
  institution = {NBER},
  type        = {Working Paper},
  number      = {12083},
  year        = {2006},
  doi         = {https://doi.org/10.3386/w12083},
}

@article{arrow2008promise,
  author  = {Arrow, Kenneth J. and Forsythe, Robert and Gorham, Michael and Hahn, Robert and Hanson, Robin and Ledyard, John O. and Levmore, Saul and Litan, Robert and Milgrom, Paul and Nelson, Forrest D. and Neumann, George R. and Ottaviani, Marco and Schelling, Thomas C. and Shiller, Robert J. and Smith, Vernon L. and Snowberg, Erik and Sunstein, Cass R. and Tetlock, Paul C. and Tetlock, Philip E. and Varian, Hal R. and Wolfers, Justin and Zitzewitz, Eric},
  title   = {The promise of prediction markets},
  journal = {Science},
  volume  = {320},
  number  = {5878},
  pages   = {877--878},
  year    = {2008},
  doi     = {https://doi.org/10.1126/science.1157679},
}

@incollection{snowberg2013prediction,
  author    = {Snowberg, Erik and Wolfers, Justin and Zitzewitz, Eric},
  title     = {Prediction markets for economic forecasting},
  booktitle = {Handbook of Economic Forecasting},
  volume    = {2},
  pages     = {657--687},
  year      = {2013},
  publisher = {Elsevier},
  doi       = {https://doi.org/10.1016/B978-0-444-53683-9.00011-6},
}

@techreport{diercks2026kalshi,
  author      = {Diercks, Anthony M. and Katz, Jared Dean and Wright, Jonathan H.},
  title       = {Kalshi and the Rise of Macro Markets},
  institution = {National Bureau of Economic Research},
  type        = {NBER Working Paper},
  number      = {34702},
  year        = {2026},
  doi         = {https://doi.org/10.3386/w34702}
}

@article{clark2007approximately,
  author  = {Clark, Todd E. and West, Kenneth D.},
  title   = {Approximately normal tests for equal predictive accuracy in nested models},
  journal = {Journal of Econometrics},
  volume  = {138},
  number  = {1},
  pages   = {291--311},
  year    = {2007},
  doi     = {https://doi.org/10.1016/j.jeconom.2006.05.023},
}

@article{corsi2009simple,
  author  = {Corsi, Fulvio},
  title   = {A simple approximate long-memory model of realized volatility},
  journal = {Journal of Financial Econometrics},
  volume  = {7},
  number  = {2},
  pages   = {174--196},
  year    = {2009},
  doi     = {https://doi.org/10.1093/jjfinec/nbp001},
}

@article{newey1987simple,
  author  = {Newey, Whitney K. and West, Kenneth D.},
  title   = {A simple, positive semi-definite, heteroskedasticity and autocorrelation consistent covariance matrix},
  journal = {Econometrica},
  volume  = {55},
  number  = {3},
  pages   = {703--708},
  year    = {1987},
  doi     = {https://doi.org/10.2307/1913610},
}

@article{benjamini1995controlling,
  author  = {Benjamini, Yoav and Hochberg, Yosef},
  title   = {Controlling the false discovery rate: A practical and powerful approach to multiple testing},
  journal = {Journal of the Royal Statistical Society: Series B},
  volume  = {57},
  number  = {1},
  pages   = {289--300},
  year    = {1995},
  doi     = {https://doi.org/10.1111/j.2517-6161.1995.tb02031.x},
}

@article{bernanke2005what,
  author  = {Bernanke, Ben S. and Kuttner, Kenneth N.},
  title   = {What explains the stock market's reaction to {Federal Reserve} policy?},
  journal = {Journal of Finance},
  volume  = {60},
  number  = {3},
  pages   = {1221--1257},
  year    = {2005},
  doi     = {https://doi.org/10.1111/j.1540-6261.2005.00760.x},
}

@article{nakamura2018high,
  author  = {Nakamura, Emi and Steinsson, J{\'o}n},
  title   = {High-frequency identification of monetary non-neutrality: The information effect},
  journal = {Quarterly Journal of Economics},
  volume  = {133},
  number  = {3},
  pages   = {1283--1330},
  year    = {2018},
  doi     = {https://doi.org/10.1093/qje/qjy004},
}

@article{gurkaynak2005sensitivity,
  author  = {G{\"u}rkaynak, Refet S. and Sack, Brian and Swanson, Eric T.},
  title   = {The sensitivity of long-term interest rates to economic news: Evidence and implications for macroeconomic models},
  journal = {American Economic Review},
  volume  = {95},
  number  = {1},
  pages   = {425--436},
  year    = {2005},
  doi     = {https://doi.org/10.1257/0002828053828446},
}

@article{harvey2016and,
  author  = {Harvey, Campbell R. and Liu, Yan and Zhu, Heqing},
  title   = {\ldots and the cross-section of expected returns},
  journal = {Review of Financial Studies},
  volume  = {29},
  number  = {1},
  pages   = {5--68},
  year    = {2016},
  doi     = {https://doi.org/10.1093/rfs/hhv059},
}

@article{white2000reality,
  author  = {White, Halbert},
  title   = {A reality check for data snooping},
  journal = {Econometrica},
  volume  = {68},
  number  = {5},
  pages   = {1097--1126},
  year    = {2000},
  doi     = {https://doi.org/10.1111/1468-0262.00152},
}

@article{demir2018does,
  author  = {Demir, Ender and Gozgor, Giray and Lau, Chi Keung Marco and Vigne, Samuel A.},
  title   = {Does economic policy uncertainty predict the {Bitcoin} returns? An empirical investigation},
  journal = {Finance Research Letters},
  volume  = {26},
  pages   = {145--149},
  year    = {2018},
  doi     = {https://doi.org/10.1016/j.frl.2018.01.005},
}

@article{liu2022risks,
  author  = {Liu, Yukun and Tsyvinski, Aleh and Wu, Xi},
  title   = {Common risk factors in cryptocurrency},
  journal = {Journal of Finance},
  volume  = {77},
  number  = {2},
  pages   = {1133--1177},
  year    = {2022},
  doi     = {https://doi.org/10.1111/jofi.13119},
}

@article{liang2022which,
  author  = {Liang, Chao and Zhang, Yaojie and Li, Xiafei and Ma, Feng},
  title   = {Which predictor is more predictive for {Bitcoin} volatility? And why?},
  journal = {International Journal of Finance \& Economics},
  volume  = {27},
  number  = {2},
  pages   = {1947--1961},
  year    = {2022},
  doi     = {https://doi.org/10.1002/ijfe.2252},
}

@article{walther2019exogenous,
  author  = {Walther, Thomas and Klein, Tony and Thu, Hien Pham and Piontek, Krzysztof},
  title   = {Exogenous drivers of {Bitcoin} and cryptocurrency volatility -- A mixed data sampling approach to forecasting},
  journal = {Journal of International Financial Markets, Institutions and Money},
  volume  = {63},
  pages   = {101133},
  year    = {2019},
  doi     = {https://doi.org/10.1016/j.intfin.2019.101133},
}

@article{corbet2018exploring,
  author  = {Corbet, Shaen and Meegan, Andrew and Larkin, Charles and Lucey, Brian and Yarovaya, Larisa},
  title   = {Exploring the dynamic relationships between cryptocurrencies and other financial assets},
  journal = {Economics Letters},
  volume  = {165},
  pages   = {28--34},
  year    = {2018},
  doi     = {https://doi.org/10.1016/j.econlet.2018.01.004},
}

@article{bouri2017hedge,
  author  = {Bouri, Elie and Moln{\'a}r, Peter and Azzi, Georges and Roubaud, David and Hagfors, Lars Ivar},
  title   = {On the hedge and safe haven properties of {Bitcoin}: Is it really more than a diversifier?},
  journal = {Finance Research Letters},
  volume  = {20},
  pages   = {192--198},
  year    = {2017},
  doi     = {https://doi.org/10.1016/j.frl.2016.09.025},
}

@article{katsiampa2017volatility,
  author  = {Katsiampa, Paraskevi},
  title   = {Volatility estimation for {Bitcoin}: A comparison of {GARCH} models},
  journal = {Economics Letters},
  volume  = {158},
  pages   = {3--6},
  year    = {2017},
  doi     = {https://doi.org/10.1016/j.econlet.2017.06.023},
}

@article{macro_crypto_volatility2025,
  title={Exploring volatility reactions in cryptocurrency markets using intraday macroeconomic news analysis},
  author={Omrane, Walid Ben and Dabbou, Halim and Saadi, Samir and Savaser, Tanseli and Sebai, Saber},
  journal={International Review of Economics \& Finance},
  volume={103},
  pages={104509},
  year={2025},
  publisher={Elsevier},
  doi={https://doi.org/10.1016/j.iref.2025.104509},
}

@article{moreira2017volatility,
  author  = {Moreira, Alan and Muir, Tyler},
  title   = {Volatility-managed portfolios},
  journal = {Journal of Finance},
  volume  = {72},
  number  = {4},
  pages   = {1611--1644},
  year    = {2017},
  doi     = {https://doi.org/10.1111/jofi.12513},
}

@article{manski2006interpreting,
  author  = {Manski, Charles F.},
  title   = {Interpreting the predictions of prediction markets},
  journal = {Economics Letters},
  volume  = {91},
  number  = {3},
  pages   = {425--429},
  year    = {2006},
  doi     = {https://doi.org/10.1016/j.econlet.2006.01.004},
}

@article{fang2020predicting,
  author  = {Fang, Tong and Su, Zhi and Yin, Libo},
  title   = {Economic fundamentals or investor perceptions? The role of uncertainty in predicting long-term cryptocurrency volatility},
  journal = {International Review of Financial Analysis},
  volume  = {71},
  pages   = {101566},
  year    = {2020},
  doi     = {https://doi.org/10.1016/j.irfa.2020.101566}
}

\newpage

\appendix

\section{Data Appendix}
\label{app:data}

\begin{table}[H]
  \centering
  \caption{Kalshi prediction market series included in the analysis.}
  \label{tab:kalshi_coverage}
  \small
  \begin{tabular}{llC{2.2cm}C{2.0cm}}
    \toprule
    Series & Macro Domain & Usable Obs. & First Active \\
    \midrule
    KXFED       & Monetary policy (rate level)  & 469 & 2023-Q1 \\
    KXCPI       & CPI inflation                 & 384 & 2023-Q1 \\
    KXCPICORE   & Core CPI inflation            & 331 & 2023-Q2 \\
    KXGDP       & Real GDP growth               & 424 & 2023-Q1 \\
    KXU3        & Unemployment rate             & 389 & 2023-Q1 \\
    KXPCECORE   & Core PCE inflation            & 294 & 2023-Q3 \\
    KXRECSSNBER & NBER recession probability    & 569 & 2023-Q1 \\
    KXACPI      & CPI level outcome (above/below threshold) & 283 & 2023-Q2 \\
    KXRATECUT   & Rate cut probability          & 28  & 2025-Q4 \\
    KXUSNFP     & Non-farm payrolls             & 0   & Never active \\
    \bottomrule
  \end{tabular}
  \par\smallskip
  \footnotesize Observations counted after one-day lag and matching with crypto returns.
\end{table}

\begin{table}[H]
  \centering
  \caption{Descriptive statistics for six cryptocurrencies, January 2023 to March 2026.}
  \label{tab:descriptive}
  \small
  \begin{tabular}{lC{1.6cm}C{1.6cm}C{1.6cm}C{1.6cm}C{1.2cm}}
    \toprule
    Asset     & Mean RVol & Std Dev & Min   & Max   & $N$   \\
    \midrule
    Bitcoin   & 0.634     & 0.194   & 0.191 & 1.341 & 1,178 \\
    Ethereum  & 0.742     & 0.254   & 0.214 & 1.731 & 1,178 \\
    Solana    & 0.930     & 0.341   & 0.186 & 2.488 & 1,178 \\
    Cardano   & 0.810     & 0.273   & 0.186 & 1.968 & 1,178 \\
    Avalanche & 0.918     & 0.309   & 0.210 & 2.101 & 1,178 \\
    Chainlink & 0.858     & 0.296   & 0.193 & 2.273 & 1,178 \\
    \bottomrule
  \end{tabular}
  \par\smallskip
  \footnotesize $\text{RVol}_{a,t}^{h=5} = \sqrt{252} \cdot \hat\sigma(r_{a,t+1},\ldots,r_{a,t+5})$ where $\hat\sigma$ is sample standard deviation.
\end{table}

\begin{table}[H]
  \centering
  \caption{All Kalshi signal coefficients for Bitcoin five-day realized volatility (model M3), sorted by $|t|$.}
  \label{tab:allsignals_btc}
  \small
  \begin{tabular}{lC{1.4cm}C{2.0cm}C{1.8cm}C{1.8cm}}
    \toprule
    Signal             & $N$  & Coefficient & HAC $t$ & $p$-value \\
    \midrule
    Fed Dovish         & 318  & $+$0.639    & $+$3.63 & 0.000$^{***}$ \\
    Recession Risk     & 389  & $-$1.536    & $-$2.76 & 0.006$^{***}$ \\
    CPI Level (KXACPI) & 195  & $+$0.412    & $+$2.40 & 0.017$^{**}$  \\
    Inflation (CPI)    & 264  & $-$0.454    & $-$1.56 & 0.119         \\
    Labor Market       & 265  & $+$0.256    & $+$1.29 & 0.196         \\
    CPI Signal         & 264  & $-$0.252    & $-$1.27 & 0.203         \\
    Monetary Policy    & 318  & $-$0.354    & $-$1.09 & 0.278         \\
    Inflation (PCE)    & 193  & $+$0.075    & $+$0.56 & 0.574         \\
    Output (GDP)       & 289  & $+$0.076    & $+$0.38 & 0.708         \\
    Inflation (Core)   & 225  & $-$0.017    & $-$0.10 & 0.919         \\
    \bottomrule
  \end{tabular}
  \par\smallskip
  \footnotesize All models include HAR components and market controls. $^{**} p < 0.05$, $^{***} p < 0.01$.
\end{table}

\end{document}